\documentstyle[11pt,newpasp,twoside,epsf]{article}
\def\edcomment#1{\iffalse\marginpar{\raggedright\sl#1\/}\else\relax\fi}
\reversemarginpar

\begin{document}
\title{SDSS Quasars and Dust Reddening}
\author{Patrick B. Hall$^1$, Philip Hopkins$^1$, Michael A. Strauss$^1$, Gordon T. Richards$^1$, \& J. Brinkmann$^2$}
\affil{$^1$~Princeton University Observatory, Princeton, NJ 08544-1001}
\affil{$^2$~Apache Point Observatory, 2001 Apache Pt. Rd., Sunspot, NM 88349}

\begin{abstract}
The Sloan Digital Sky Survey can be used to detect and characterize red and
reddened quasars.  In Richards et al.\ (2003), we showed that 
6\% of SDSS quasars have red colors consistent with significant dust reddening
by an extinction curve similar to that of the Small Magellanic Cloud (SMC).  
We estimate that a further 10\% of the luminous quasar population is missing 
from our magnitude-limited SDSS sample.
More recent work (Hopkins et al., in preparation) confirms the conclusion
that dust reddening is the primary explanation for SDSS quasars
in the red tail of the color distribution.
Fitting orthogonal first- and second-order polynomials to SDSS quasar photometry
measures the slope and curvature of each object's UV/optical spectrum.
The slope vs. curvature distribution is elongated along the axis predicted for
SMC-like reddening, while the axes predicted for LMC- or MW-like reddening
provide significantly poorer fits.  Extension to longer wavelengths using a
smaller sample of SDSS/2MASS matches confirms this result
at high significance.
\end{abstract}

\section{Introduction}

The SDSS Quasar Survey is very well
suited to the detection and characterization of moderately reddened quasars.
Quasar candidates are selected in the $i$-band, which is less affected by dust
extinction than traditional $B$-band selection.
It is large ($\sim$30,000 quasars to date) and relatively deep ($i\leq19.1$
and $i\leq20.2$ for low- and high-redshift quasar candidates, respectively).
It includes high-quality ($\sigma_{\rm phot}\leq0.05$), five-band photometric data
for all quasar candidates plus high-quality, moderately high-resolution
($R\sim2000$) spectra for most of them.  
In this contribution we highlight some results of our initial study of
reddened quasars (Richards et al.\ 2003), as well as some more recent results
from work in progress.

\section{Using Relative Colors to Identify Red and Reddened Quasars}

The left-hand panels in Figure 1 show the observed colors of SDSS quasars as a
function of redshift in $u$$-$$g$, $i$$-$$z$, and $g$$-$$i$.
Calling a quasar `red' based on an observed color cut (e.g., $u$$-$$g$$>$$0.6$,
upper left) is subject to redshift-dependent systematic errors.
Instead, we subtract the median quasar color at $z$$<$$2.2$ and plot the
resulting histograms of {\em relative} colors $\Delta$$($$u$$-$$g$$)$, 
$\Delta$$($$i$$-$$z$$)$, and $\Delta$$($$g$$-$$i$$)$ in the right-hand panels
in Figure 1.  The $\Delta$$($$i$$-$$z$$)$ histogram (middle right)
is nearly Gaussian, but shorter wavelength colors such as $\Delta$$($$u$$-$$g)$
(top right) have increasingly prominent red tails.  Intrinsic red power laws
would produce identical histograms in each color, while synchrotron emission
cutoffs would produce redder colors at longer wavelengths.  Therefore, 
dust reddening must be responsible for the red tails in the histograms.

Figure~2a~shows~the~distribution~of~$\Delta$($g$$-$$i$)~vs.~$z$~for~$\sim$15,800~SDSS~quasars.
Solid lines show the effect of SMC reddening with the indicated
$E$$($$B$$-$$V$$)$ as a function of $z$ on a quasar with intrinsic
$\Delta$($g$$-$$i$)=0.2.  Unambiguously dust-reddened quasars lie to the right
of the leftmost line.  From them, we estimate that our sample misses only
10\% of the quasar population due to dust extinction with $E(B-V)\leq0.28$.
Figure 2b shows composite quasar spectra as a function of $\Delta(g-i)$;
the moderately dust-reddened composite is made from quasars lying
between the darker lines in Figure 2.  As expected, each dust-reddened
composite has significant curvature --- a power-law connecting continuum
windows at 1450 and 4000\,\AA\ (dotted line) overpredicts the observed flux
longward of 5000\,\AA.

\section{Spectral Slope and Curvature at Ultraviolet-Optical Wavelengths}

\begin{figure}
\plotfiddle{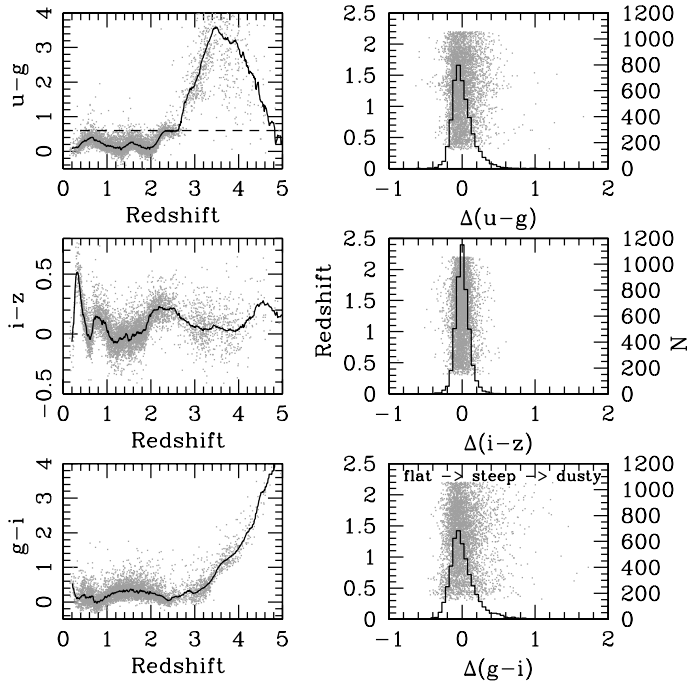}{3.25in}{0.0}{47.5}{47.5}{-150}{-80}
\caption{{\em Left:} Observed (dots) and median (line) quasar colors
as a function of redshift.  {\em Right:} Relative quasar colors after 
subtraction of the median color as a function of redshift.  The histograms
show that shorter-wavelength color distributions have stronger red tails.}
\vspace{-0.1cm}
\end{figure}

\begin{figure} 
\plotone{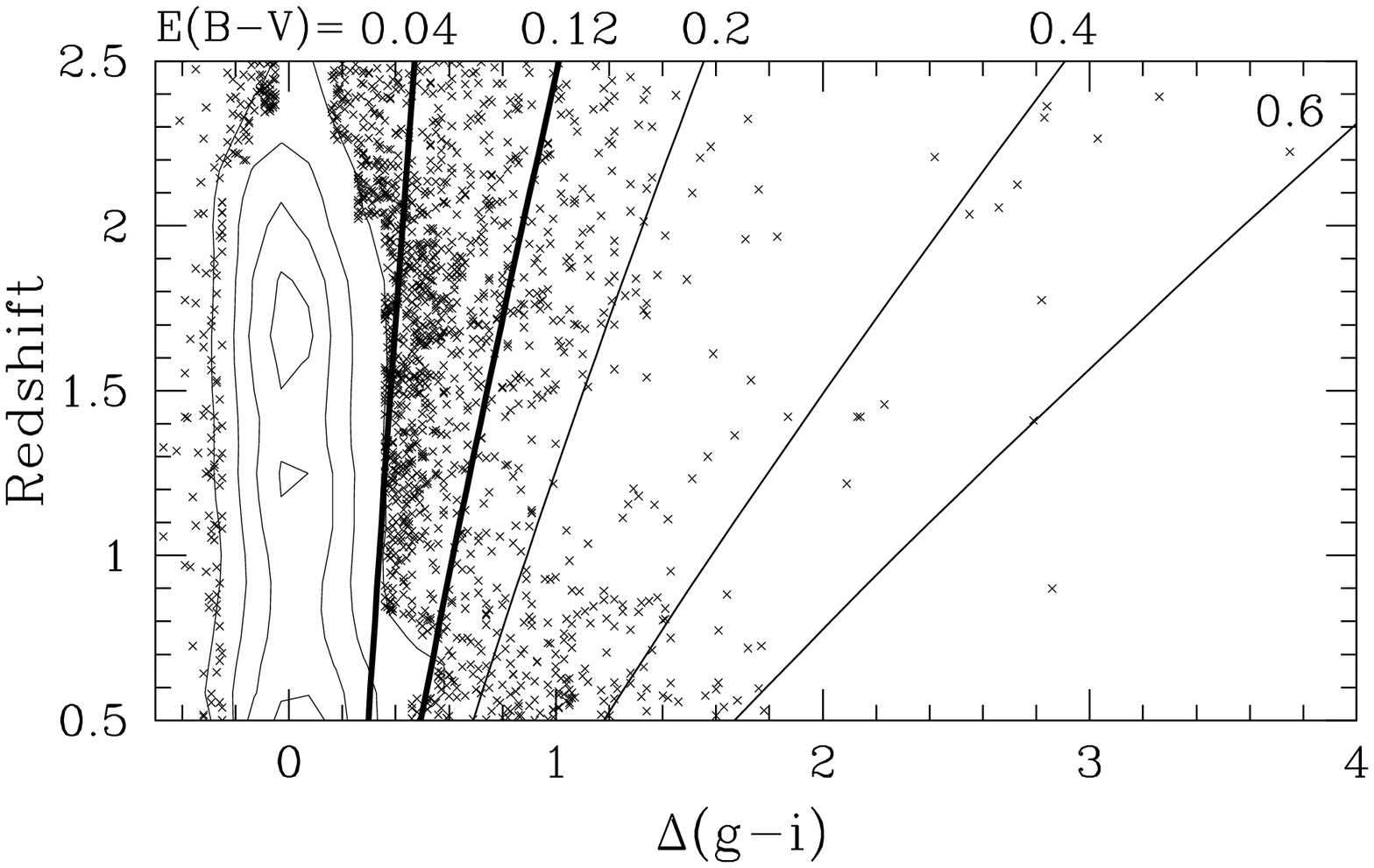}\\ \plotone{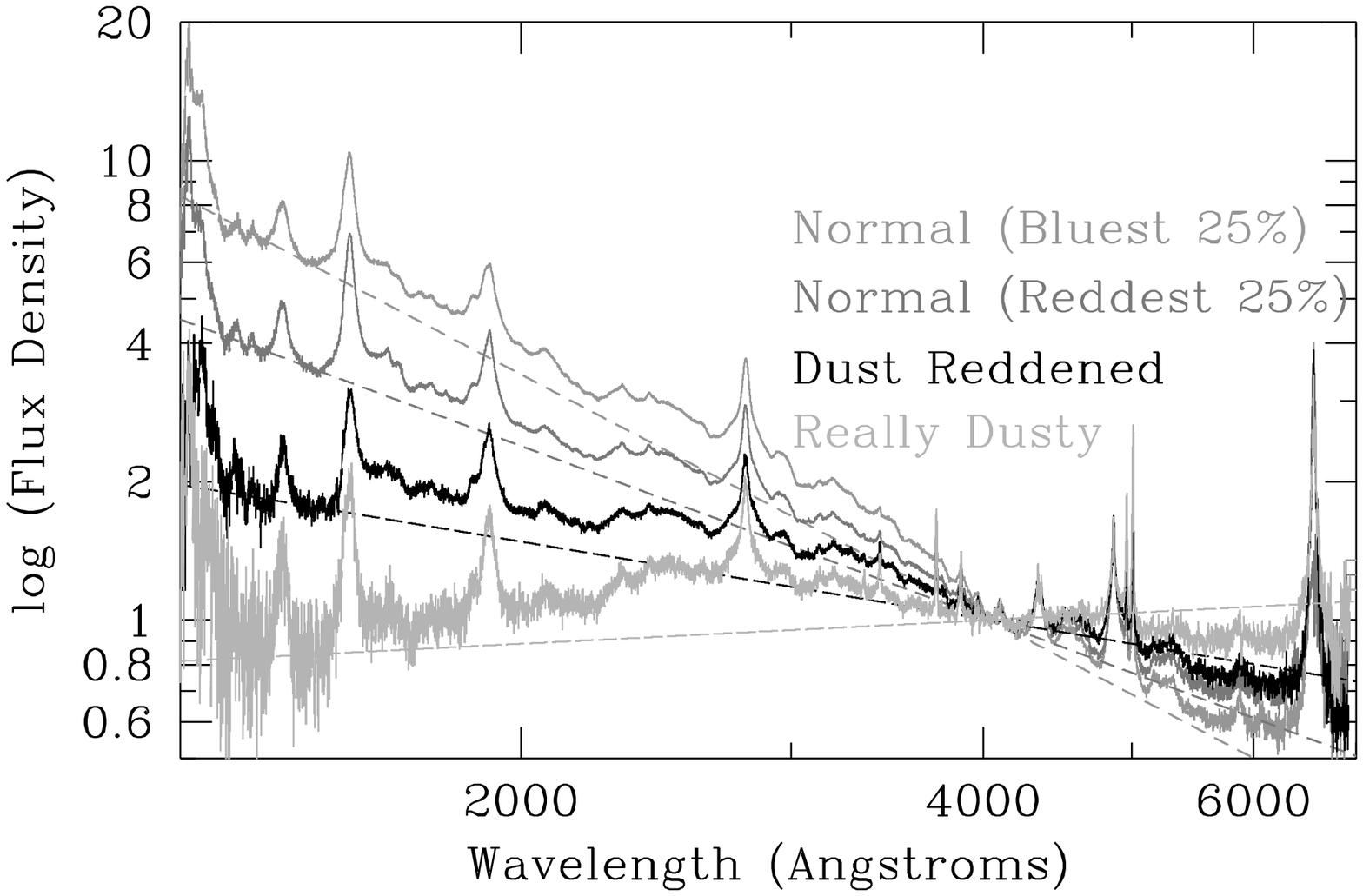}
\caption{{\em Top ($a$):}
Distribution of SDSS quasars in relative color $\Delta$($g$$-$$i$)
versus redshift.  Contours show normal quasars, distributed between 
$-$0.3$<$$\Delta$($g$$-$$i$)$<$0.3 due to the dispersion in
power-law spectral indices.  Outside this range (points), the
excess of redder quasars over bluer ones is due to dust reddening (plus
host galaxy contamination at $z$$<$0.9).  Solid lines show the effects of dust
reddening with the indicated SMC extinction curve $E$$($$B$$-$$V$$)$.
{\em Bottom ($b$):} Composite spectra as a function of $\Delta$($g$$-$$i$) color,
plus attempted fits to underlying power-law continua.}
\end{figure}

In Hopkins et al.\ (in prep.), we fit orthogonal first- and second-order
polynomials to the relative magnitudes (see above) of $\sim$15,800 SDSS quasars.
This procedure yields the `slope' and `curvature' of each object's `photometric
spectrum' via the coefficients $c^*_1$ and $c^*_2$ of the fitted
first- and second-order Chebyshev polynomials.
We~fit~the~photometry~because~it~has~higher~S/N,~better~calibration,~and~a~longer
wavelength baseline than the spectra.  We extend the wavelength baseline by a
factor of $\sim$3, at the expense of larger photometric uncertainties,
using a subsample of $\sim$2,600 SDSS quasars also detected by
the Two-Micron All-Sky Survey.

Figure 3 shows that the slope vs. curvature distribution for matched SDSS-2MASS
quasars (contours and squares) is elongated
along the axis predicted for SMC-like reddening (plus signs and line),
while the relations predicted for the LMC (diamonds) or MW
(crosses) extinction curves provide significantly poorer fits.
SMC-like extinction is more common than LMC-like extinction
along our lines of sight to quasars, and MW-like extinction
is rare along such sightlines.

\begin{figure}
\plotfiddle{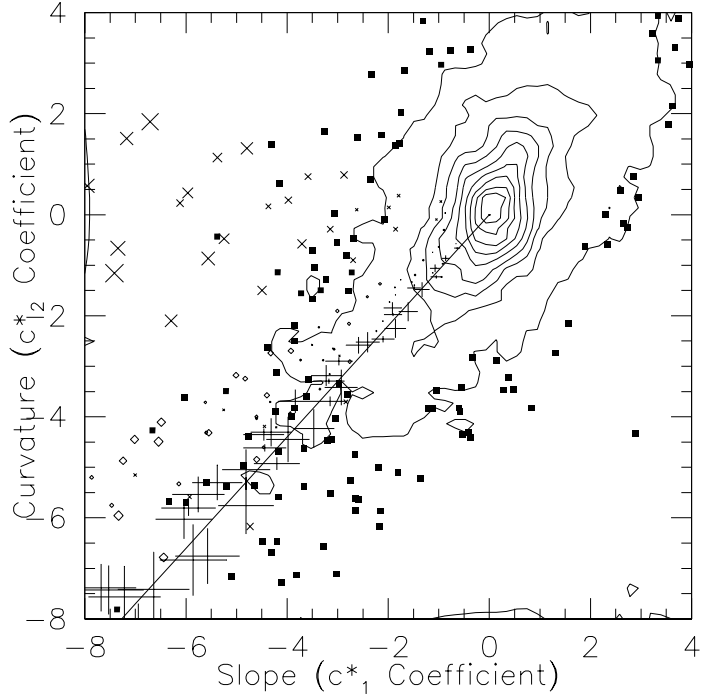}{3.45in}{0.0}{47.5}{47.5}{-145}{-70}
\caption{The slope vs. curvature distribution (see text) for matched SDSS-2MASS
quasars (contours and squares).  Slope and curvature values are also shown for
simulated quasars, over a range of redshifts, reddened by the extinction curves
of the SMC (plus signs and solid line), the LMC (diamonds), and the Milky Way 
(crosses).}
\vspace{-0.25cm}
\end{figure}

\acknowledgments
Funding for the creation and distribution of the SDSS Archive
(http://www.sdss.org/) has been provided by the Alfred P. Sloan
Foundation, the Participating Institutions, the National Aeronautics
and Space Administration, the National Science Foundation, the
U.S. Department of Energy, the Japanese Monbukagakusho, and the Max
Planck Society.


\end{document}